\documentclass[conference]{IEEEtran}
\IEEEoverridecommandlockouts
\usepackage{cite}
\usepackage{amsmath,amssymb,amsfonts}
\usepackage{algorithmic}
\usepackage{graphicx}
\usepackage{textcomp}
\usepackage{xcolor}
\usepackage{siunitx}
\usepackage{subfigure}
\usepackage{multirow}
\def\BibTeX{{\rm B\kern-.05em{\sc i\kern-.025em b}\kern-.08em
    T\kern-.1667em\lower.7ex\hbox{E}\kern-.125emX}}
\begin{document}

\title{NB-IoT Uplink Synchronization by Change Point Detection of Phase Series in NTNs\\
\thanks{This paper is supported in part by National Natural Science Foundation of China Program(62271316, 62101322), National Key R\&D Project of China (2018YFB1802201), Shanghai Key Laboratory of Digital Media Processing (STCSM 18DZ2270700) and the Fundamental Research Funds for the Central Universities. }

\thanks{The authors Jiaqi Jiang, Yihang Huang, Yin Xu, Runnan Liu, Xiaowu Ou and Dazhi He are with School of Electronic Information and Electrical Engineering, Shanghai Jiao Tong University. The corresponding author is Yin Xu (E-mail: xuyin@sjtu.edu.cn).}}
\author{Jiaqi Jiang, Yihang Huang, Yin Xu, Runnan Liu, XiaoWu Ou and Dazhi He\\
School of Electronic Information and Electrical Engineering, Shanghai Jiao Tong University, Shanghai, China\\
\{jiangjiaqi1999, huangyihangsg, xuyin, liurunnan,xiaowu\_ou,hedazhi\}@sjtu.edu.cn
}
\maketitle

\begin{abstract}
Non-Terrestrial Networks (NTNs) are widely recognized as a potential solution to achieve ubiquitous connections of Narrow Bandwidth Internet of Things (NB-IoT). In order to adopt NTNs in NB-IoT, one of the main challenges is the uplink synchronization of Narrowband Physical Random Access procedure which refers to the estimation of time of arrival (ToA) and carrier frequency offset (CFO). Due to the large propagation delay and Doppler shift in NTNs, traditional estimation methods for Terrestrial Networks (TNs) can not be applied in NTNs directly. In this context, we design a two stage ToA and CFO estimation scheme including coarse estimation and fine estimation based on abrupt change point detection (CPD) of phase series with machine learning. Our method achieves high estimation accuracy of ToA and CFO under the low signal-noise ratio (SNR) and large Doppler shift conditions and extends the estimation range without enhancing Random Access preambles.
\end{abstract}


\section{Introduction}
Narrow Bandwidth Internet of Things (NB-IoT) is a key motivation for Beyond fifth Generation (B5G) mobile networks which in turn will support applications of massive Machine Type of Communication (mMTC) such as smart cities, mobile health, and other large-scale NB-IoT use cases  \cite{b1}. With the rapid growth in the amount of IoT and NB-IoT devices, especially for devices in remote areas where terrestrial IoT can not cover, Non-terrestrial networks (NTNs) are introduced to help achieve seamless global coverage. In NTNs, Low and very Low Earth Orbit (LEO and vLEO) satellite are widely recognized as the most suitable way to achieve ubiquitous IoT coverage with its lower power consumption and propagation delay compared with Geostationary Earth Orbit (GEO) satellite. The Third Generation Partnership Project (3GPP) has standardized the NB-IoT in release-16 \cite{b2} to enable communications between IoT devices and has started a study item centered on NB-IoT/eMTC support for NTNs to adopt the NTNs in NB-IoT \cite{b3}. As is well known, Random Access (RA) procedure is widely used to establish a wireless link and enable the data transmission between the user equipments (UEs) and satellite base station. In this context, one of the main challenges is the NB-IoT uplink synchronization for RA in NTNs which refers to accurate time of arrival (ToA) and carrier frequency offset (CFO) estimation in particular. In NTNs, due to the impairments caused by large propagation delay and Doppler shift, the original estimation method \cite{b4,b5,b6,b7} designed for TNs does not apply in NTNs. In \cite{b4}, Brute Force (BF) algorithm based on differential correlation is proposed to detect ToA while its accuracy degrades with the presence of residual CFO. \cite{b5} exploits frequency hopping rules in RA to eliminate the impact of CFO when estimating ToA but its estimation range is limited by preamble format. Many existed studies \cite{b8,b9,b10,b11} adopt Global Navigation Satellite System (GNSS) in user equipments to pre-compensate the large propagation delay and Doppler shift and then tackle the issue with small residual ToA and CFO after pre-compensation. 
Furthermore, \cite{b11} uses cumulative sum algorithm (CUSUM) to detect the change point of wavelet to estimate the ToA and CFO but its performance heavily depends on how well the actual data follows the assumed distribution. However, the assumption of GNSS-assisted devices with high accuracy positioning for NB-IoT in NTNs is questionable. Major challenges include \cite{b12,b13}:
\begin{enumerate}
\item GNSS receiver needs to frequently compensate CFO and ToA which is not suitable for NB-IoT devices with limited computation capability and power consumption.
\item Natural radio propagation impairments and building blockage can undermine the positioning performance of GNSS.
\item The time synchronization algorithm in presence of GNSS is vulnerable to different types of intentional attacks.
\end{enumerate}
Such impacts heavily interfere the pre-compensation performance of GNSS. Hence the residual ToA and CFO may exceed the estimation range, leading to overall performance degradation. Consequently, instead of using GNSS, we firstly use a system-level method proposed in \cite{b14} to compensate the large CFO to around 600Hz which relies on the information obtained from downlink synchronization. Then, aiming to eliminate the impact of large ToA and achieve the accurate estimation of residual ToA and CFO, we propose a novel method based on the change point detection (CPD) of phase series to identify the existence of preamble signal. We firstly multiply the received signal by local RA preamble and then extract its phase series. It can be explored that the phase series represent periodicity when preamble signal exists and follow random distribution otherwise. The periodicity reflects CFO and the location of each period reflects ToA. We find that the slope of phase series changes abruptly at the junction between each period, caused by phase ambiguity. Then, motivated by \cite{b18}, we use autoenconders with a time-invariant representation (TIRE) to detect the change points. The reason to select TIRE method relies on its accurate detection ability of the slope change despite the data is not generated from parametric probability distribution. Besides, the sensitivity of detection performance to common noise can be mitigated during the process of feature learning by autoencoders. The coarse estimation of ToA is conducted by first two change points of phase series. Then CFO is estimated from the period and the fine estimation of TOA is calculated based on the CFO and the location of first change point. 

The remainder of this paper is summarized as follows. Section \uppercase\expandafter{\romannumeral2} describes the RA structures and the preamble signals. Section \uppercase\expandafter{\romannumeral3} proposes our ToA and CFO estimation method based on phase series. In Section \uppercase\expandafter{\romannumeral4}, the simulation results are presented and the conclusion of this paper is in Section \uppercase\expandafter{\romannumeral5}.

\section{SYSTEM MODEL}
 This work focuses on addressing the accurate estimation of ToA and CFO in RA procedure via satellite, which manages the uplink synchronization in NTNs. We consider D2 scenario among six reference scenarios presented by 3GPP in \cite{b15}. It means our system is operated in S-band at a carrier-frequency of \SI{2}{G\hertz}, deployed at the altitude of 600km. In this section, the basic structures of RA preamble are introduced and then we model the transmission signal and present our assumptions.
\subsection{Random Access Preamble Structure}
\begin{figure}[htbp]
	\centering
	\includegraphics[width=0.45\textwidth]{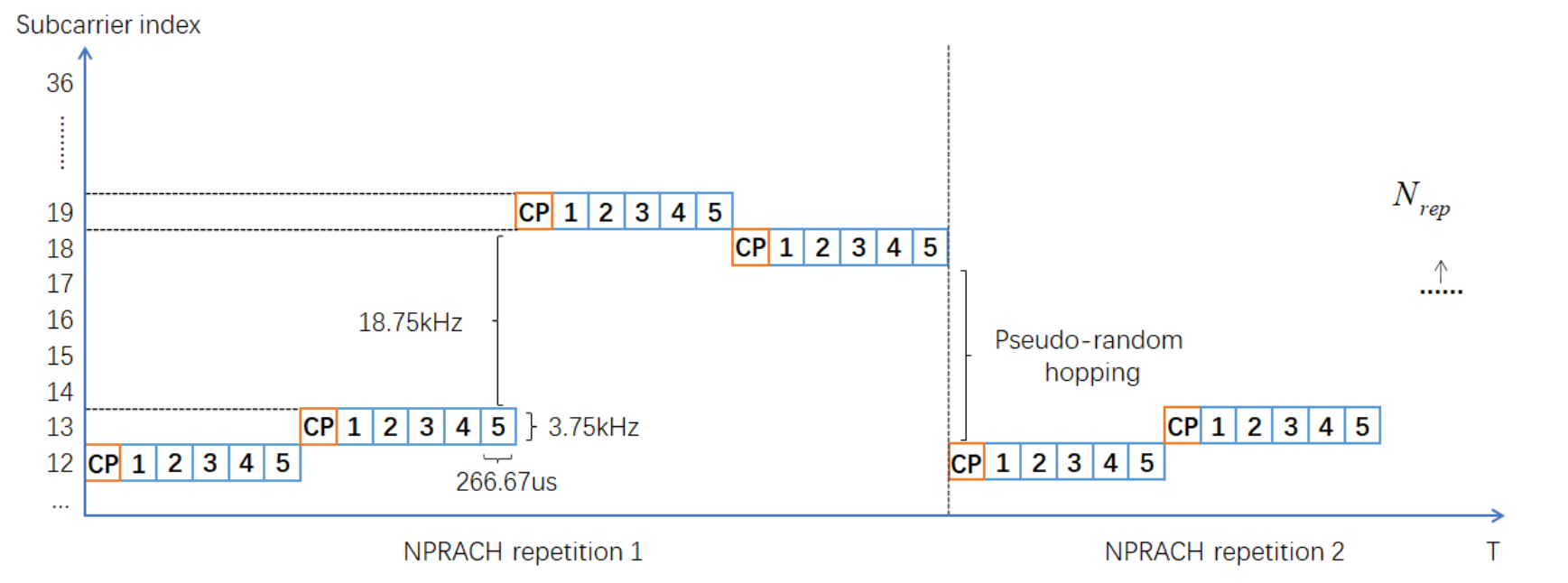}
	\caption{Preamble structure in Format 1}
	\label{fig1}
\end{figure}

In NB-IoT, the RA preamble is transmitted on Narrowband Physical Random Access Channel (NPRACH) which occupies a bandwidth of \SI{180}{k\hertz}. There are three commonly used formats tagged by format 0 to format 2 for RA preamble in FDD mode. The starting time of NPRACH transmission is set by $N_{start}$. Four symbol groups constitute a basic unit preamble and the number of preamble units in whole RA procedure is determined by $N_{rep}$. One symbol group (SG) consists of one cyclic prefix (CP) plus five identical symbols. In format 0, the length of CP                                                                                                                                                                                       $(66.67\mu s)$ is a quarter of the length of one basic symbol $(266.67\mu s)$ and in format 1 both lengths are identical. In frequency domain, each SG occupies one subcarrier with the bandwidth of 3.75KHz and the total number of subcarriers is denoted as $N_{sc}$. The  initial subcarrier for NPRACH is defined by $N_{off}$. The frequency spacing between SGs follows a predefined hopping pattern and the frequency spacing from one unit preamble to another follows a pseudo-random selection principle summarized in \cite{b16}. In fact, the allocation of frequency resources in one unit preamble only depends on the subcarrier location of first SG. Fig. \ref{fig1} shows an example of preamble structure. 
\subsection{NPRACH Signal Transmission}
The baseband signal transmitted by NPRACH can be expressed as
\begin{equation}\label{eq1}
	s_{m,p}(n)=\sum_k^{N-1}S_{m,p}[k]e^{j2\pi\frac{k}{N}n}
\end{equation}
where $s_{m,p}(n)$ denotes the $n$-th sample of the time domain waveform in $p$-th symbol of the $m$-th SG. $S_{m,p}[k]$ denotes the $m$-th symbol on the $p$-th subcarrier during the $m$-th SG. The range of sample point $n$ belongs to $[N_{m,p}-N_{CP},N_{m,p}+N-1]$, in which $p\in[0,L-1]$. $L$ denotes the number of symbols in one SG and $N_{m,p}=mN_g+pN$, with $N_g=N_{CP}+pN$ being the size of one SG, $N_{CP}$ denoting the length of $CP$ and $N$ denoting the length of one symbol. Based on \cite{b16}, $S_{m,p}[k]=\begin{cases}
1&k=N_{sc}(m)\\
0&others
\end{cases}$, $N_{sc}(m)$ is the subcarrier index occupied by $m$-th SG, so the transmitted signal can be rewritten as
\begin{equation}\label{eq2}
	s_{m,p}(n)=\sum_m^{N_{sym}}S_{m,p}[N_{sc}(m)]e^{j2\pi\frac{N_{sc}(m)}{N}n}
\end{equation}
$N_{sym}=4\times N_{rep}$ represents the number of SG in format 1. After transmitting to the receiver on NPRACH, the $n$-th sample of $p$-th symbol of the received signal can be written as
\begin{equation}\label{eq3} 
	y_{m,p}(n)=h_me^{j2\pi f_{off}(n-D)}s_{m,p}(n-D)+w_{m,p}(n)
\end{equation}
where $f_{off}$ denotes the CFO normalized by sampling frequency and $D$ denotes the ToA normalized by the symbol duration. $w_{m,p}(n)$ is the noise term and $h_m$ is the channel coefficient for the $m$-th SG. Here we consider the impact of Doppler rate. The changing rate of the frequency offset over time changes from $f_{off}$ to $f_{off}+\alpha (n-D)$ with the presence of Doppler rate $\alpha$ normalized by squared sampling frequency. In this context, the received signal can be changed as
\begin{equation}\label{eq4}
 	y_{m,p}(n)=h_me^{j2\pi [f_{off}(n-D)+\frac{1}{2}\alpha (n-D)^2]}s_{m,p}(n-D)+w_{m,p}(n) 
\end{equation}
The following assumptions are made based on the considered scenario: \romannumeral1) the terrestrial UEs are directly connected to LEO satellite. \romannumeral2) CFO remains constant during the RA procedure. \romannumeral3) The Doppler rate of each geographical location can be estimated by LEO satellite. For the sake of simplicity, two kinds of transmission channels concluding Additive White Gaussian Noise (AWGN) channel and Tapped Delay Line-C (TDL-C) channel are considered in the following analysis.
\section{estimation method based on change point detection}
\begin{figure}[htbp]
	\centering
	\includegraphics[width=0.45\textwidth]{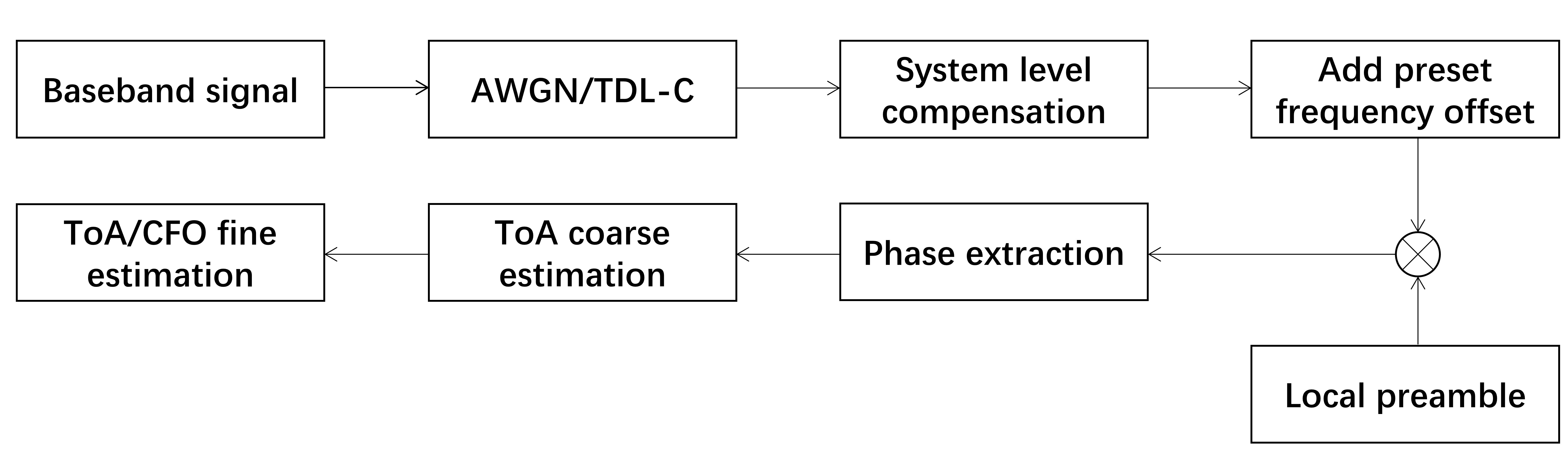}
	\caption{Estimation process}
	\label{fig2}
\end{figure}

Based on the preamble index attained from the preamble detection, the same baseband preamble signal can be selected at the local receiver side. Fig. \ref{fig2} illustrates the overall estimation scheme. A system level solution based on the information of initial downlink synchronization is used to reduce the large CFO to a relatively small scale \cite{b14}. In most cases, the residual frequency offset is about 600Hz \cite{b17} which is still much larger than estimation range of existed studies \cite{b4,b5,b6,b7,b8,b9,b10,b11}. Therefore, we consider the frequency uncertainty as 600Hz in the following analysis. 
\subsection{Phase Series Analysis}\label{AA}
We first multiply the received signal by conjugation of local baseband signal:
\begin{align}\label{eq5}
	r(n)&=y_{m,p}(n)\times s_{m,p}^\ast(n)\notag\\
	    &=h_me^{j2\pi [f_{off}(n-D)+\frac{1}{2}\alpha (n-D)^2]}\times\notag\\
	    &e^{j{-2\pi\frac{N_{sc}(m)}{N}D}}+w_{m,p}(n)
\end{align}
The phase series extracted from $r(n)$ is expressed as
\begin{align}\label{eq6}
	\varphi (n)&=\varPhi (y_{m,p}(n)\times s_{m,p}^\ast(n))\notag\\
	&=2\pi[f_{off}(n-D)+\frac{1}{2}\alpha (n-D)^2]-2\pi\frac{N_{sc}(m)}{N}D\notag\\
	&+\Delta_h+\Delta_w
\end{align}
$\Delta_h$ denotes the phase of channel coefficient and $\Delta_w$ denotes the phase of noise. On AWGN channel, $\Delta_h=0$. On TDL-C channel, $\Delta_h$ represents the micro phase fluctuation. Because the Doppler rate varies from $0$ to \SI{-620}{\hertz} in our scenario \cite{b15}, the parameter $\alpha$ normalized by squared sampling frequency is far less than $f_{off}$ normalized by sampling frequency. So $\alpha$ can be neglected when phase series is analyzed. The impact of $\Delta_w$ to phase series can be weakened by smoothening the mean of received signal. Then the phase series can be simplified as
\begin{equation}\label{eq7}
	\varphi (n)=2\pi f_{off}(n-D)-2\pi\frac{N_{sc}(m)}{N}D
\end{equation}
It can be seen that the phase series is a linear function of time sample $n$ with the slope as $2\pi f_{off}$ and intercept as $-2\pi D(f_{off}+\frac{N_{sc}(m)}{N})$, manifested as periodic change of phase series in the range of $[-\pi,\pi]$.

\subsection{ToA and CFO Estimation Scheme}
\begin{figure}[htbp]
	\centering
	\begin{minipage}[t]{0.3\textwidth}
		\centering
	\includegraphics[width=2in]{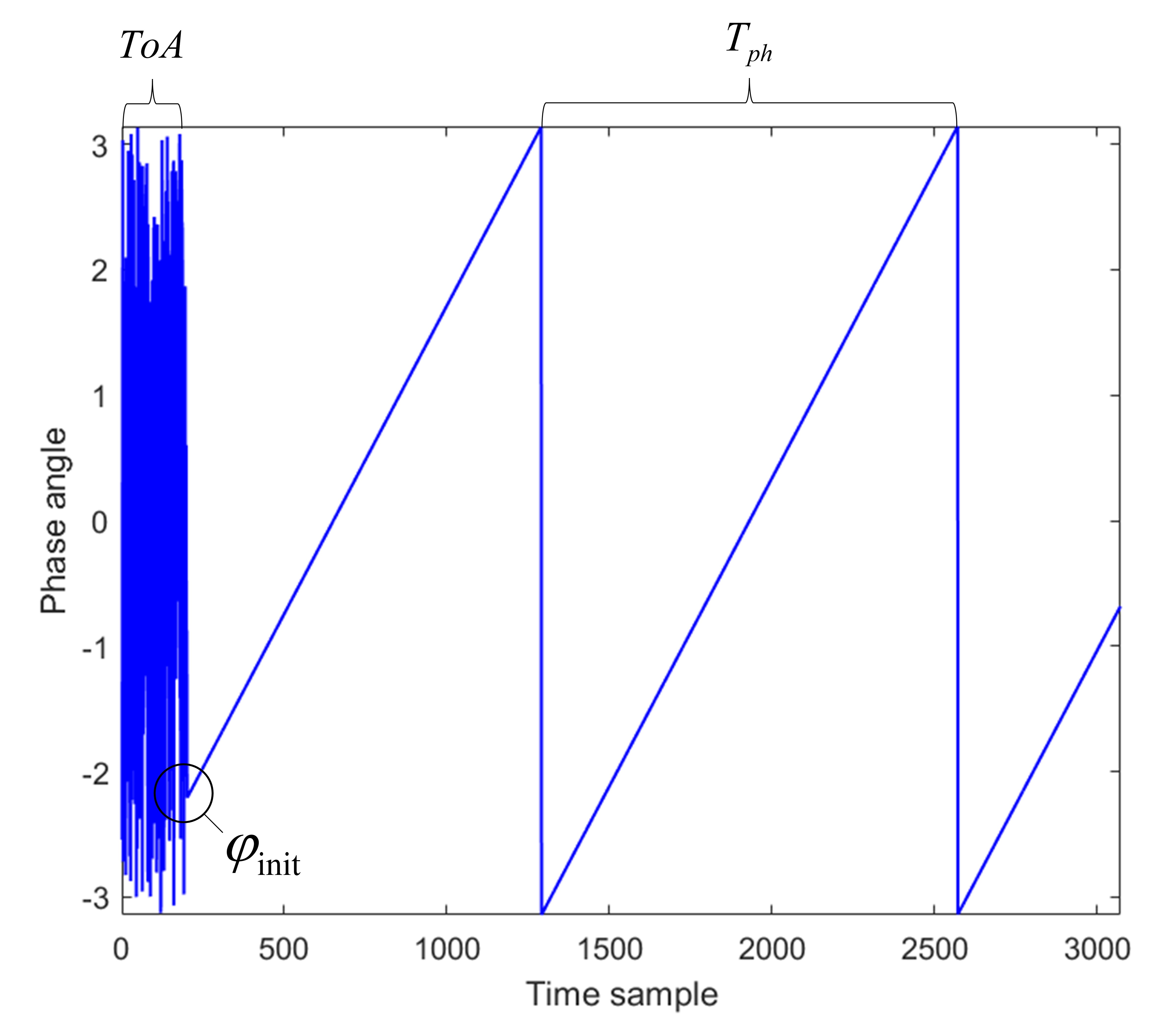}
	\caption{Phase series with CFO=\SI{1500}{\hertz}}
	\label{fig3}
\end{minipage}
\end{figure}

Due to the range of residual CFO after system level compensation being [\SI{-600}{\hertz},\SI{600}{\hertz}], the linearity of phase series would be undermined in extreme cases. In order to transfer the linear feature to the periodic change of phase series, a fixed frequency offset can be added to received signal after the system level compensation. Here we set a \SI{1000}{\hertz} offset plus positive CFO and \SI{-600}{\hertz} plus negative CFO to enhance the estimation performance. The frequency uncertainty then turns to [(-1600,-1000),(1000,1600)]\SI{}{\hertz} which means there exists a constant prior deviation between the estimation value and the true value. During the NPRACH process, based on the assumptions \romannumeral2), the slope of phase series remains constant and the intercept changes with the SG. Fig. \ref{fig3} shows the phase series with the ToA of $200$ sample points and the CFO of \SI{1500}{\hertz}.

\begin{figure}[htbp]
	\centering
	\begin{minipage}[t]{0.4\textwidth}
	\centering
	\includegraphics[width=2.5in]{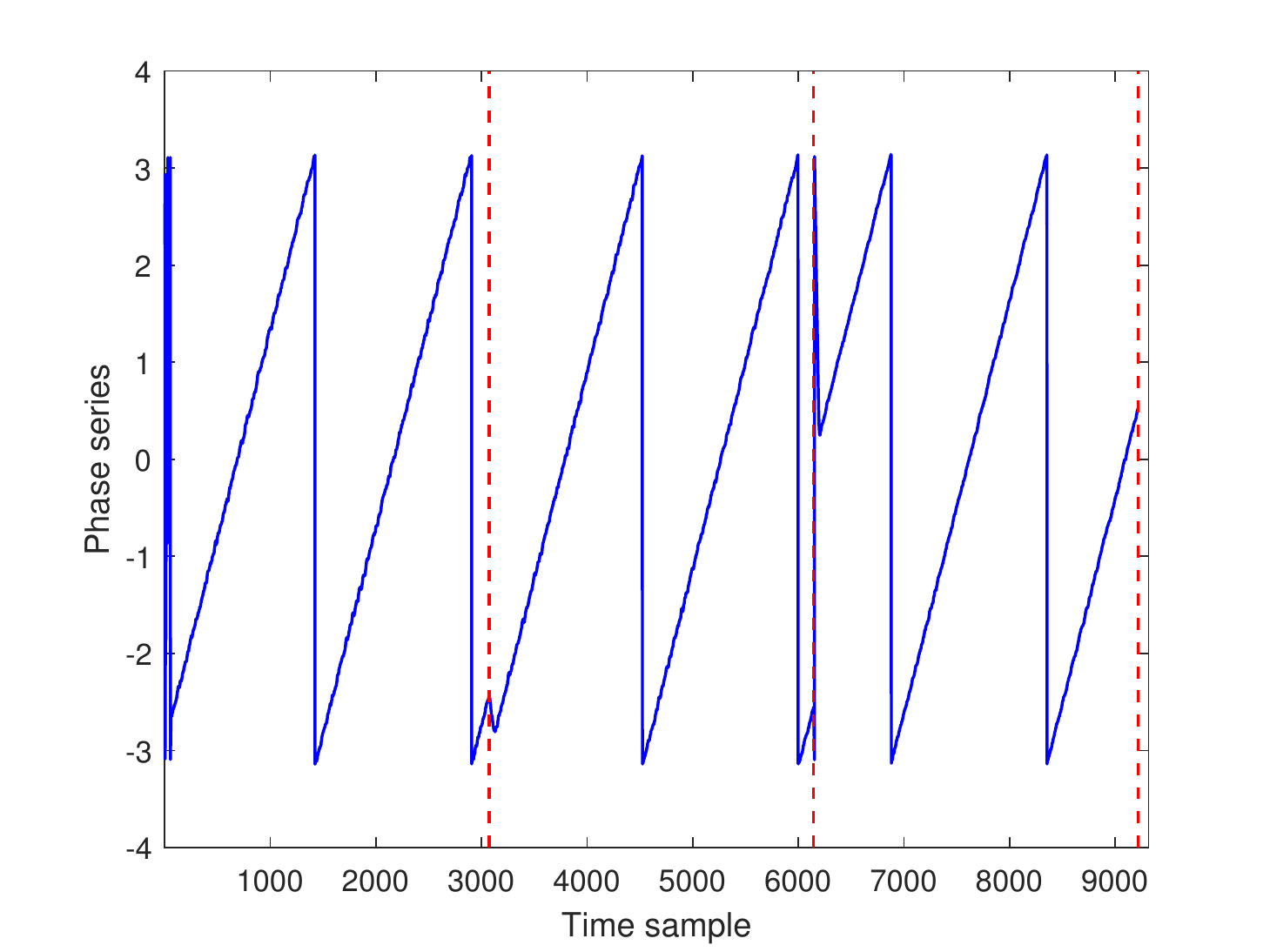}
	\caption{Segmentation of phase series}
	\label{fig4}
    \end{minipage}
\end{figure}

Next, we reveal that the information of ToA and CFO can be reflected in phase series. Determined by configuration parameters and formulation (\ref{eq7}), the initial phase at the start of NPRACH transmission can be expressed as
\begin{align}\label{eq8}
	\varphi_{int}&=\varphi(N_{start})\notag\\
	&=2\pi[f_{off}(N_{start}-D)]-2\pi\frac{N_{off}}{N}N_{start}
\end{align}
Then the ToA and CFO can be calculated as
\begin{align}\label{eq9}
	&D=n_l-\frac{|\pm\pi-\varphi_{int}|}{f_{off}}\\
	&f_{off}=\frac{1}{T_{ph}}
\end{align}
$T_{ph}$ denotes the period of phase series and $n_l$ denotes the timing location of the first period and the sign of $\pi$ depends on the sign of CFO. Now the estimation of ToA and CFO refers to the estimation of $T_{ph}$ and $n_l$. We use the change point detection method which would be introduced in following part to determine $T_{ph}$ and $n_l$. Due to the 2$\pi$ phase ambiguity, the solution to equation (\ref{eq9}) is not unique, resulting several ToA candidates to be selected. A discrimination method based on Doppler rate \cite{b14} is utilized to select the true ToA among candidates. Because both the Doppler rate and ToA are related to the position of UE, the precise positioning information is not required. To illustrate, assuming the true ToA is $104.7\mu s$ and the solutions to formulation (\ref{eq9}) are $104.7\mu s$, $371.3\mu s$, $638.0\mu s$ with corresponding Doppler rate of \SI{-297}{\hertz/s}, \SI{-252}{\hertz/s}, \SI{-215}{\hertz/s} respectively. If the Doppler estimation is \SI{-240}{\hertz/s}, then correct ToA selection is $371.3\mu s$ because its Doppler rate is the nearest one. Then in ToA coarse estimation process, the distance of first two detected change points is recorded as $T_{ph}$. Coarse ToA values which are used to compensate large ToA can be calculated based on the $T_{ph}$. Noting that the coarse $T_{ph}$ estimation would result in wrong ToA candidates selection, leading major ToA estimation error in coarse estimation process.
After coarse estimation of ToA, the accurate CFO is estimated based on the whole preamble signal. In order to eliminate the impact of different intercept of phase series in each SGs, we divide the phase series into segments with the length of one symbol group and then calculate the distances in each segments, which is shown in Fig. \ref{fig4}. The estimation of CFO relies on the average distances of change points in each segments and we use the Tree Sigma Guidelines as the postprocessing scheme to exclude anomalous values of distances to increase estimation accuracy. Then fine estimation of ToA can be conducted.
\subsection{Slope Change Detection Method}
\begin{figure}[htbp]
	\centering
	\includegraphics[width=0.4\textwidth]{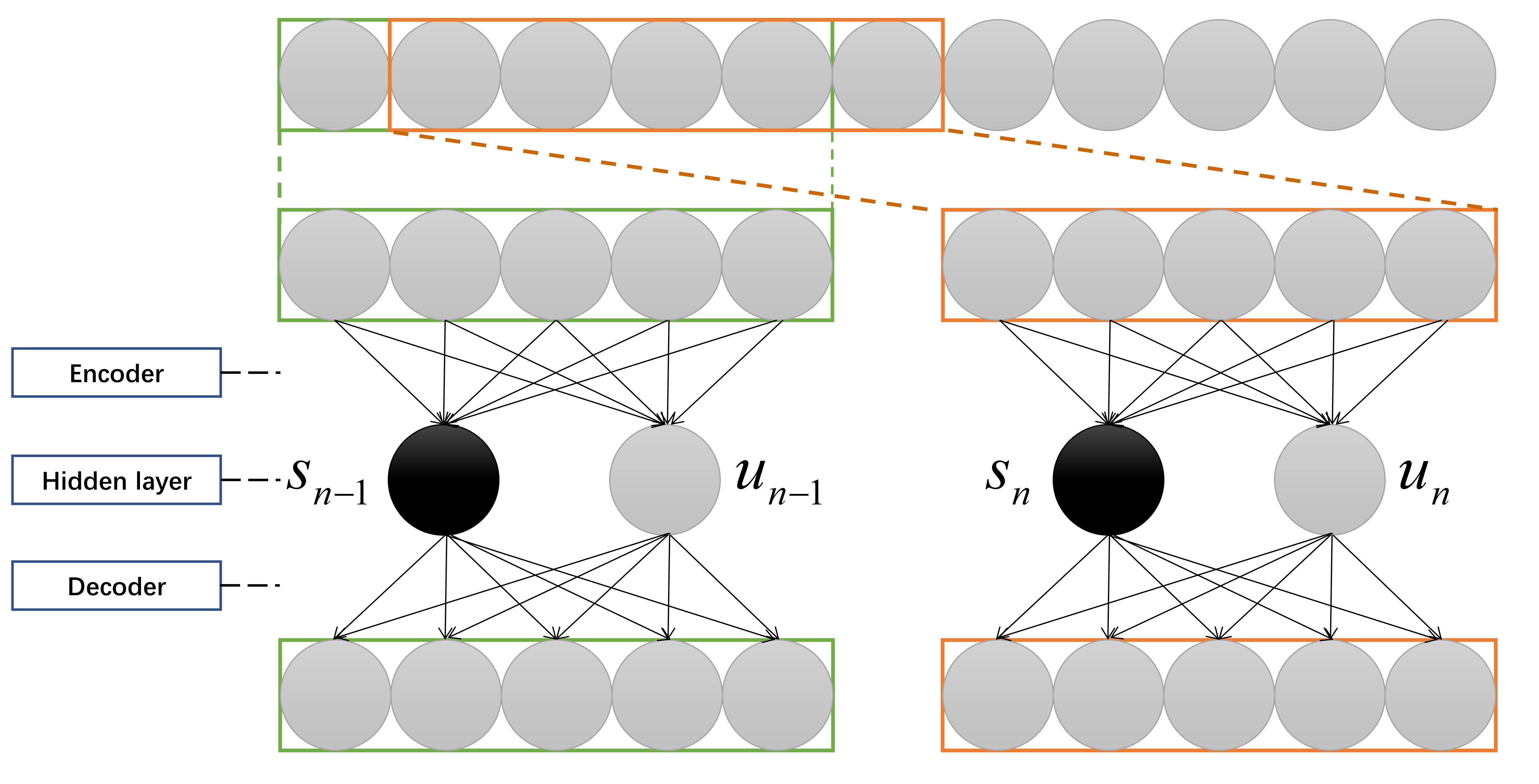}
	\caption{TIRE structure}
	\label{fig5}
\end{figure}

We use autoencoders with a time-invariant representation (TIRE) to detect slope change points of phase series and then calculate the $T_{ph}$ and $n_l$ \cite{b18}.  We first segment phase series into consecutive windows with the length of N. The autoencoders are utilized to extract features of hidden layer from the consecutive windows. These features can be divided into time-invariant features $s_n$ and instantaneous features $u_n$ (\ref{eq10}). The structure is shown in Fig. \ref{fig5}. 
\begin{equation}
	\label{eq10}
	h_n=[(s_n)^T,(u_n)^T]^T
\end{equation}
In (\ref{eq10}) $h_n$ denotes the encoded output from encoder layer of the $n$-th window. Invariant features refer to the statistical characteristics that change only when change point exists in consecutive windows. That means the differences of time-invariant features between time windows can manifest the abrupt change. The differences are summarized by the defined dissimilarity measure $D$ 
\begin{equation}
	\label{eq11}
	D_n=||s_n^{TD}-s_{n+N}^{TD}||_2
\end{equation}
$N$ denotes the time-domain window size. The change points are located at the peaks of dissimilarity measure $D_n$. In order to reduce the alarming rate, we exploit the prominence of peaks to determine the location of change points by comparing with a predifined threshold.

\begin{figure}[htbp]
	\centering
	\begin{minipage}[t]{0.5\textwidth}
	\centering
	\includegraphics[width=3.5in]{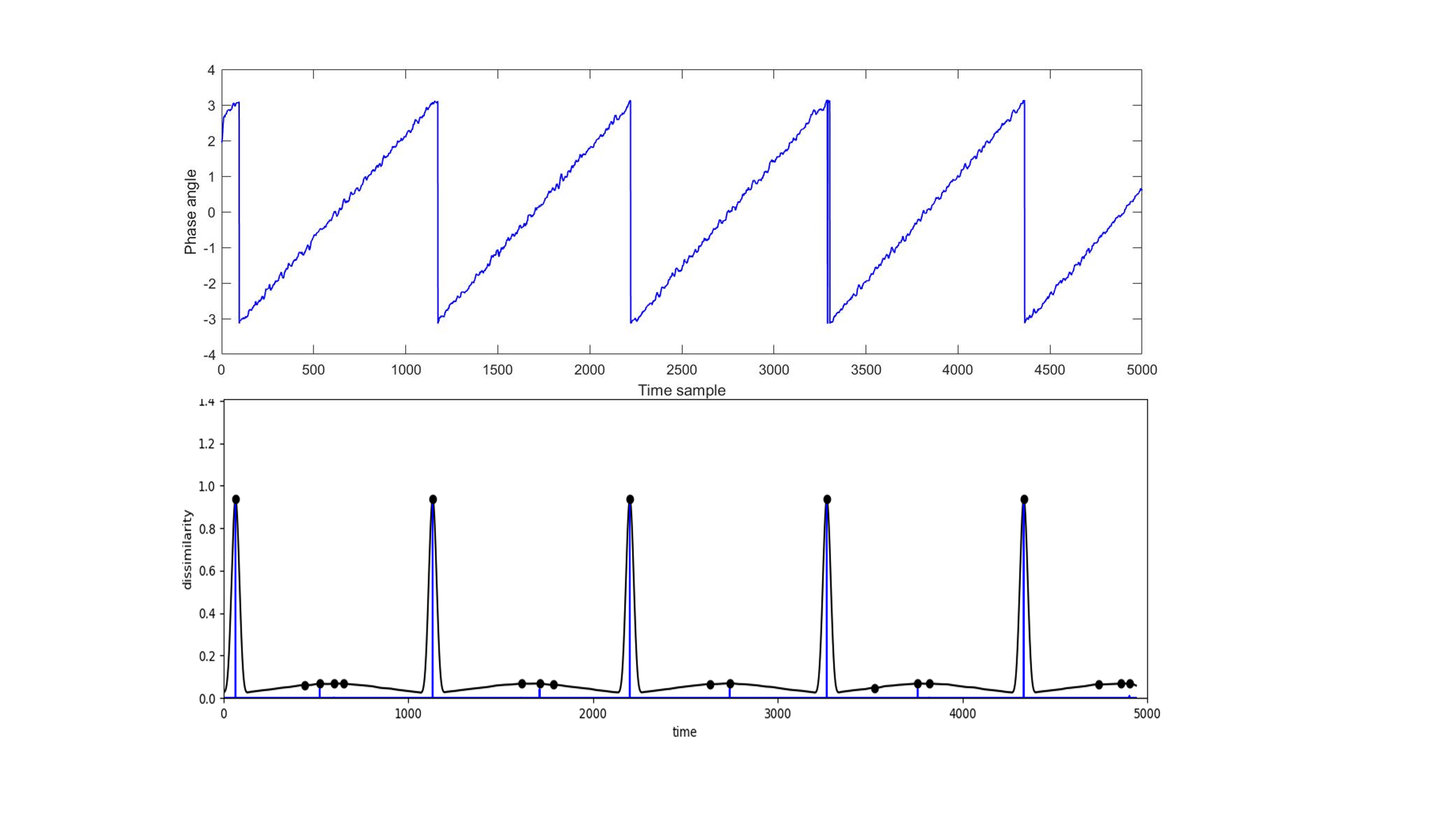}
	\caption{Correspondence of phase series and dissimilarity measure}
	\label{fig6}
\end{minipage}
\end{figure}

In our context, we set the number of time invariant features as one in time domain which refers to the slope of phase series. Fig. \ref{fig6} demonstrates the phase series and its corresponding dissimilarity measure. The blue lines represent the prominence of peaks. Change points can be detected when the prominence exceeds predefined threshold. 
\section{simulation results}
In this section, we first discuss the coarse estimation performance of  ToA to determine the range of residual ToA through Monte Carlo simulations. Then residual ToA and CFO estimation results are presented. The simulation parameters are listed in TABLE \ref{tab2}. 
\begin{table}[htbp]
	\caption{simulation parameters}
	\begin{center}
		\begin{tabular}{|c|c|}
		\hline
		\textbf {Parameters} &\textbf{Setting}\\
		\hline
		Sampling rate $(f_{s})$&1.92 Mbps\\
		\hline
		Subcarrier spacing $(\Delta f)$&3.75 kHz\\
		\hline
		Satellite altitude&600 km\\
		\hline
		Beam diameter&90 km\\
		\hline
		Elevation angle&$90^\circ$\\
		\hline
		FFT length $(N)$&512\\
		\hline
		CP length $(N_{CP})$&$266.7 \mu s$\\
		\hline
		Frequency uncertainty&$[-600,600]Hz$\\
		\hline
		\end{tabular}
		\label{tab2}
	\end{center}
\end{table}
\subsection{ToA Coarse Estimation Performance}
The normalization of ToA and CFO is done on the symbol duration and sampling frequency. Table \ref{tab3} reports the ToA coarse estimation performance in terms of absolute average estimation error and max estimation error. Based on the estimation performance, the range of residual ToA is set as $[-100\mu s,100\mu s]$ in the following simulation.
\begin{table}[htbp]
	\caption{ToA coarse estimation performance}
	\begin{center}
		\begin{tabular}{|c|c|c|}
			\hline
			\textbf {SNR} &\textbf{Average error($\mu s$)}&\textbf{Max error($\mu s$)}\\
			\hline
			SNR=3&28.1&41.1\\
			\hline
			SNR=0&52.7&61.5\\
			\hline
			SNR=-3&76.9&88.5\\
			\hline
		\end{tabular}
		\label{tab3}
	\end{center}
\end{table}
\subsection{CFO Estimation Performance}
\begin{figure}[htbp]
	\centering
	\begin{minipage}[t]{0.4\textwidth}
		\centering
		\includegraphics[width=2.5in]{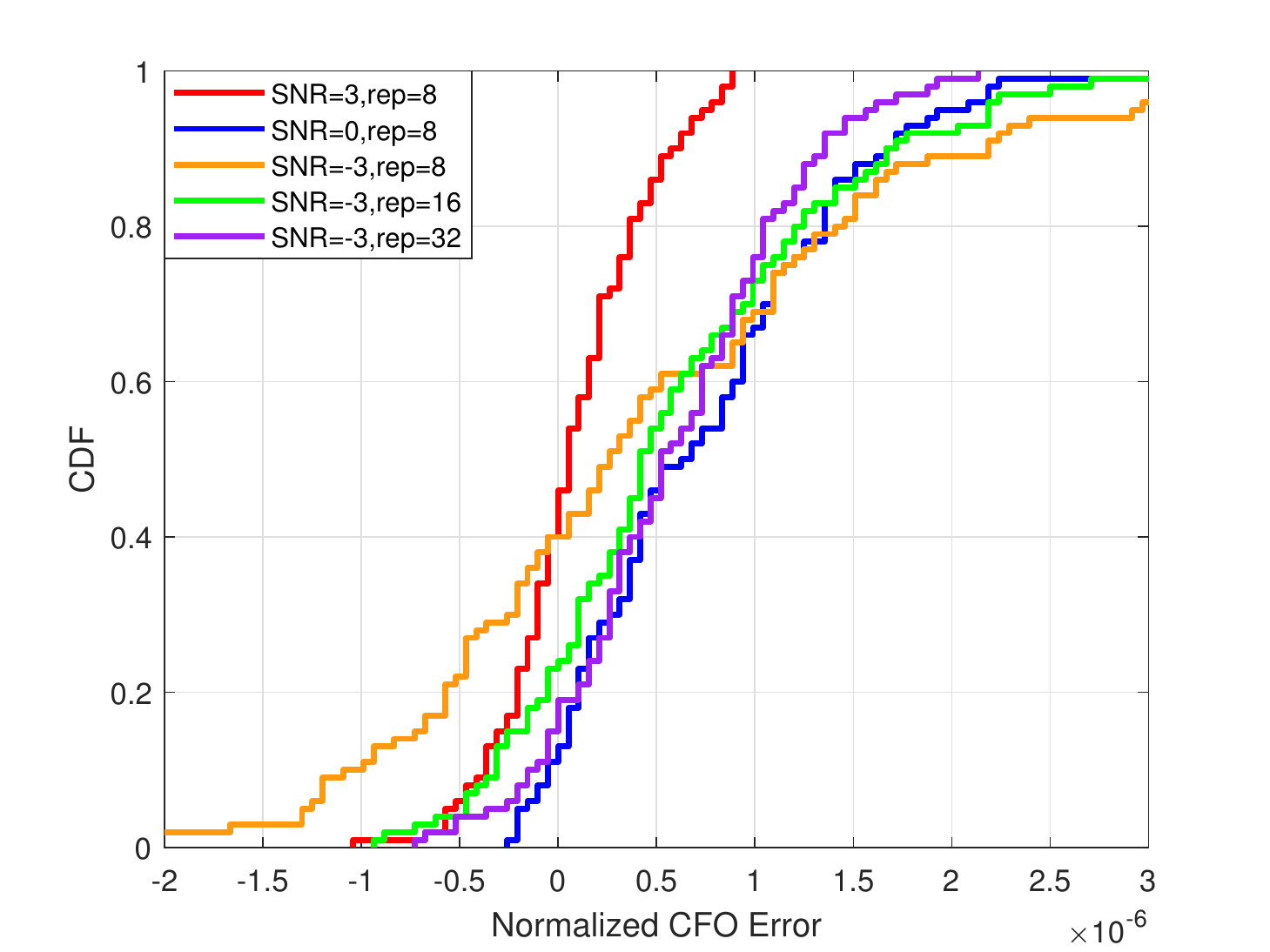}
		\caption{CDF of normalized CFO estimation error}
		\label{fig10}
	\end{minipage}
\end{figure}

Fig. \ref{fig10} shows the cumulative distribution function (CDF) of normalized CFO error with different SNR (SNR=3dB, SNR=0dB, SNR=-3dB) when the $N_{rep}$ is set as 8 and the CDF of normalized CFO error with different $N_{rep}$ ($N_{rep}=8, N_{rep}=16, N_{rep}=32$) when SNR is set as -3dB. It can be found that the absolute CFO estimation error in 99\% cases is less than \SI{1.92}{\hertz} with SNR=3dB. However, the estimation error grows when decreasing the SNR. This can be solved by increasing the repetition of basic preamble units $(N_{rep})$.  When $N_{rep}=32$, the max absolute CFO error is less than \SI{4.5}{\hertz}. The estimation accuracy of CFO grows with the $N_{rep}$ resulting from the amount of detected change points in phase series. The amount also increases with the growth of CFO. Hence, the estimation performance with large CFO can be guaranteed.
\subsection{ToA Fine Estimation Performance}

We simulate the ToA estimation process on two kinds of channels: AWGN channel and TDL-C channel. Here the technique proposed in \cite{b11} and the Brute Force (BF) algorithm based on differential correlation \cite{b4} are compared with our method. In \cite{b11}, the author exploits Stationary Discrete Wavelet Transform (S-DWT) to decompose the received signal into 8 levels. It can be found that the decomposed sequence $y_i$ follows two different distributions before and after time delay $\tau$. Then the hypothesis of no change point in whole sequence is described as $H_0:\theta =\theta_0 for 1\leq i\leq N$ and if one change point appears at $i=\tau$, the hypothesis is presented as $H_1:\theta =\theta_0 for 1\leq i\leq \tau,\theta=\theta_1 for \tau \leq i\leq N$. Based on cumulative sum (CUSUM) algorithm of change point detection, the log-likelihood ratio (LLR) of hypothesis reaches the maximum after the change. This algorithm is indicated as DWT in the following. In \cite{b4}, the BF algorithm based on differential correlation detects the peak of cross correlation values. In the following, Fig. 8-9 report the normalized ToA error with different SNR (SNR=3dB, SNR=0dB, SNR=-3dB) when the $N_{rep}$ is set as 8 on AWGN channel and TDL-C channel. 
It can be seen that the proposed method outperforms the DWT and differential correlation method both on AWGN channel and on TDL-C channel due to the steepest slope of CDF. The performance of differential correlation method is  heavily ruined because of the presence of noise. The performance of DWT heavily depends on how well the actual data follows the assumed Gaussian distribution which can be proved by the simulation results that the estimation error of DWT increases apparently on TDL-C channel.  Since we use the autoencoders with TIRE to detect the change points of phase series without advanced assumption of distribution, our method is applicable for various channels besides AWGN channel. When the SNR decreases from 3dB to -3dB on AWGN channel and TDL-C channel, the curves follow the same trend. And the max absolute normalized ToA error changes from 2.1$\mu s$ to 4.2$\mu s$ on AWGN channel and from 2.7$\mu s$ to 5.3$\mu s$ on TDL-C channel which demonstrates the robustness against the noise. 
The whole estimation process is conducted in the time domain and both ToA and CFO are estimated in one process of change point detection in this method, which conserves the computational complexity of Discrete Fourier Transform (DFT). The structure of autoencoders with TIRE in this context can be pre-trained using the training data from actual preamble signals to speed up the uplink synchronization. In addition, our method enlarges the estimation range of CFO 
compared with existed methods. Thus, the overall performance proves that our method is fully suitable to address the uplink synchronization for NB-IoT in NTNs. 
\begin{figure*}[htbp]
	\centering
	\subfigure[]{
			\begin{minipage}[t]{0.3\textwidth}
					\centering
					\includegraphics[width=2in]{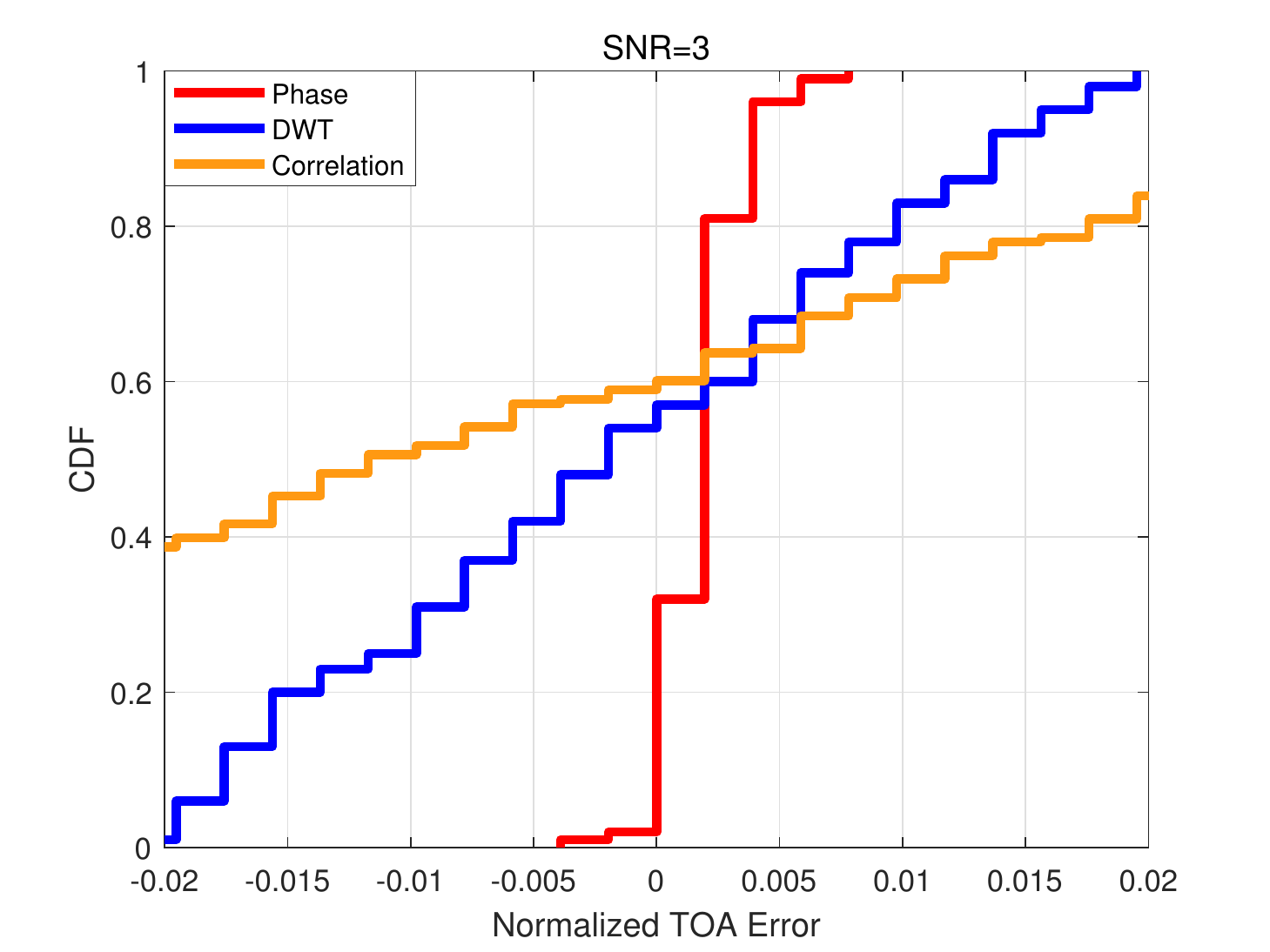}
				\end{minipage}%
		}%
	\subfigure[]{
			\begin{minipage}[t]{0.3\textwidth}
					\centering
					\includegraphics[width=2in]{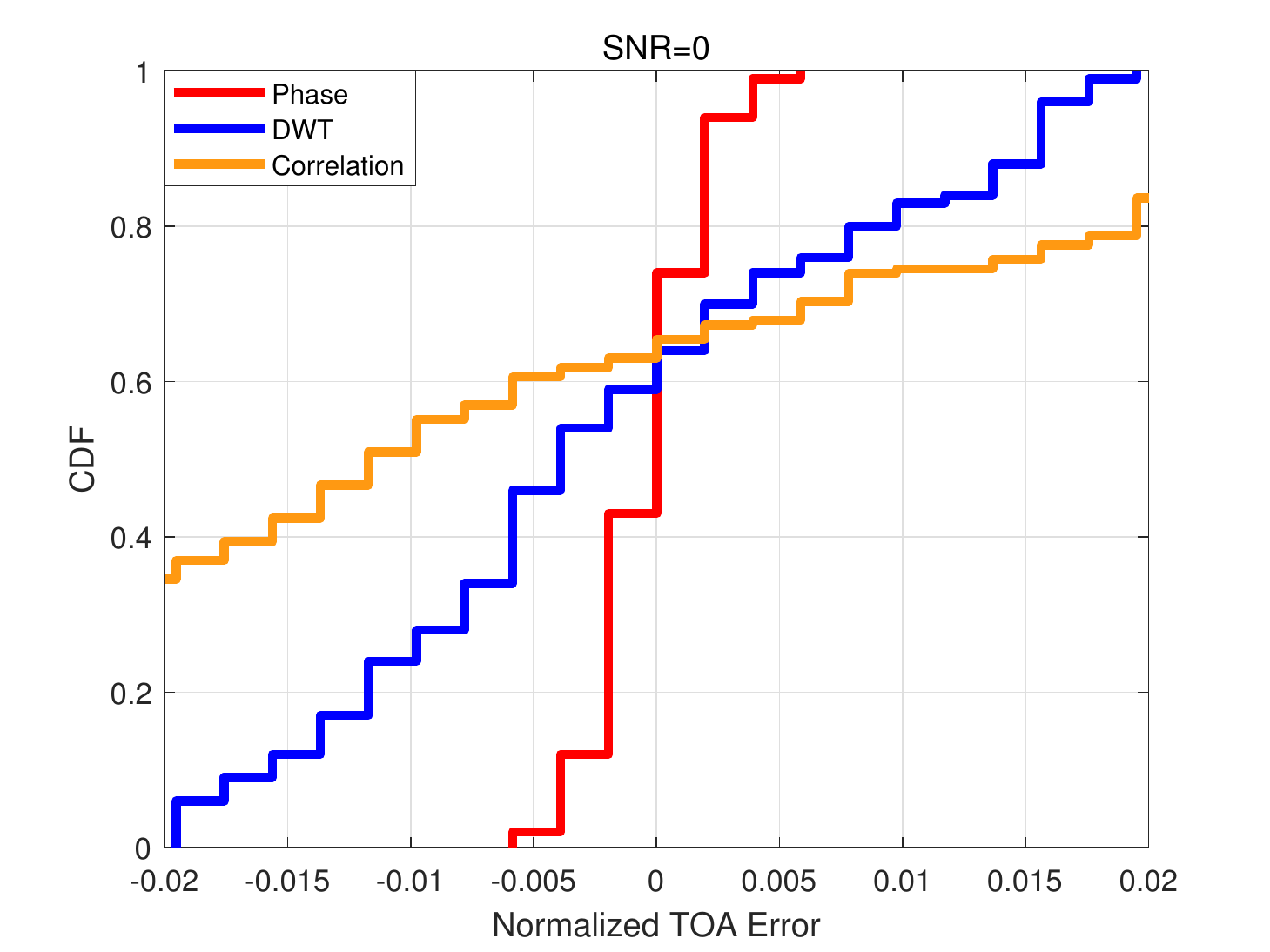}
				\end{minipage}%
		}%
	\subfigure[]{
		\begin{minipage}[t]{0.3\textwidth}
				\centering
				\includegraphics[width=2in]{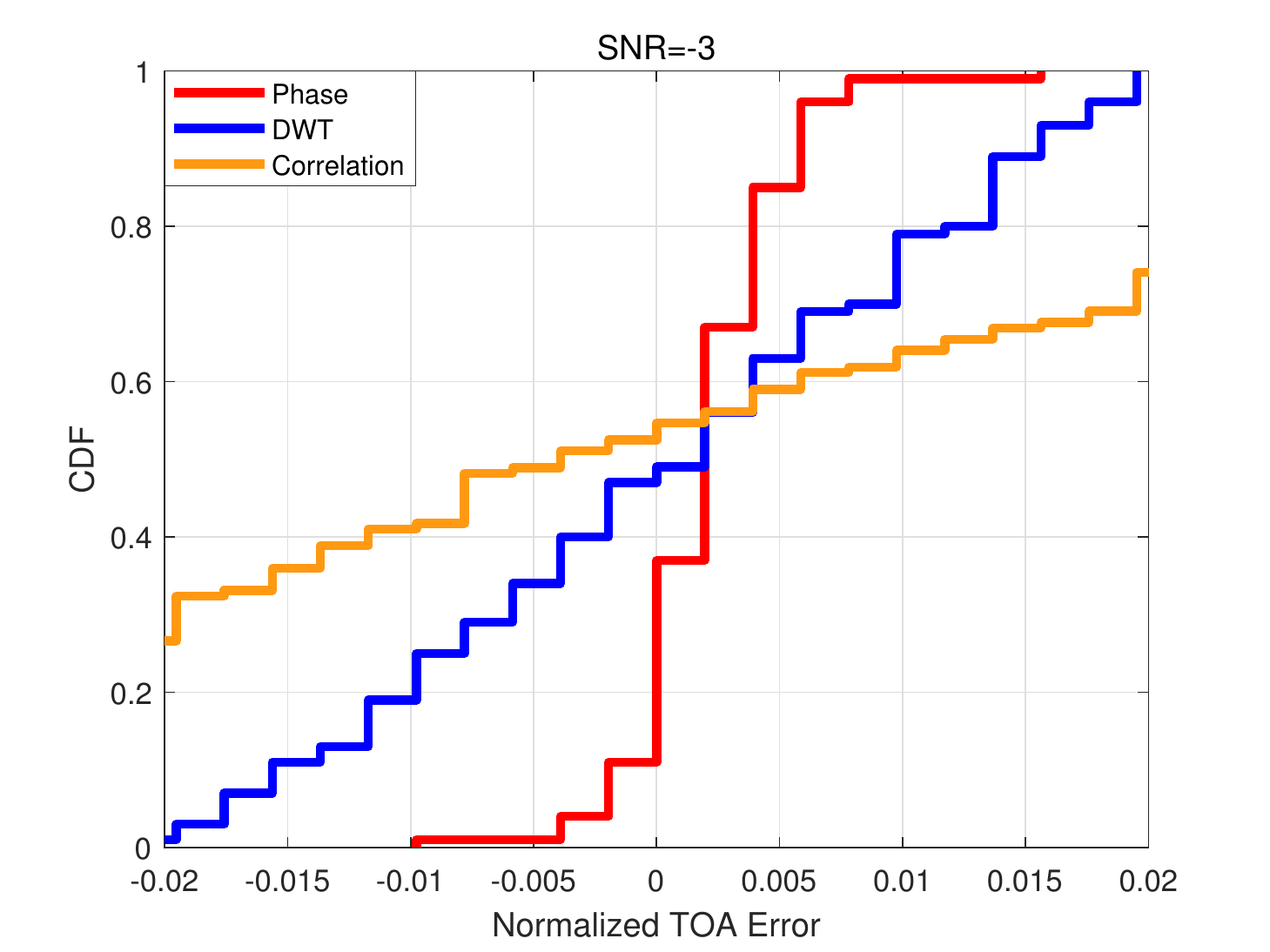}
			\end{minipage}%
	}%

	\centering
	\caption{CDF of normalized ToA estimation error on AWGN channel}
\end{figure*}
\begin{figure*}[htbp]
	\centering
	\subfigure[]{
			\begin{minipage}[t]{0.3\textwidth}
					\centering
					\includegraphics[width=2in]{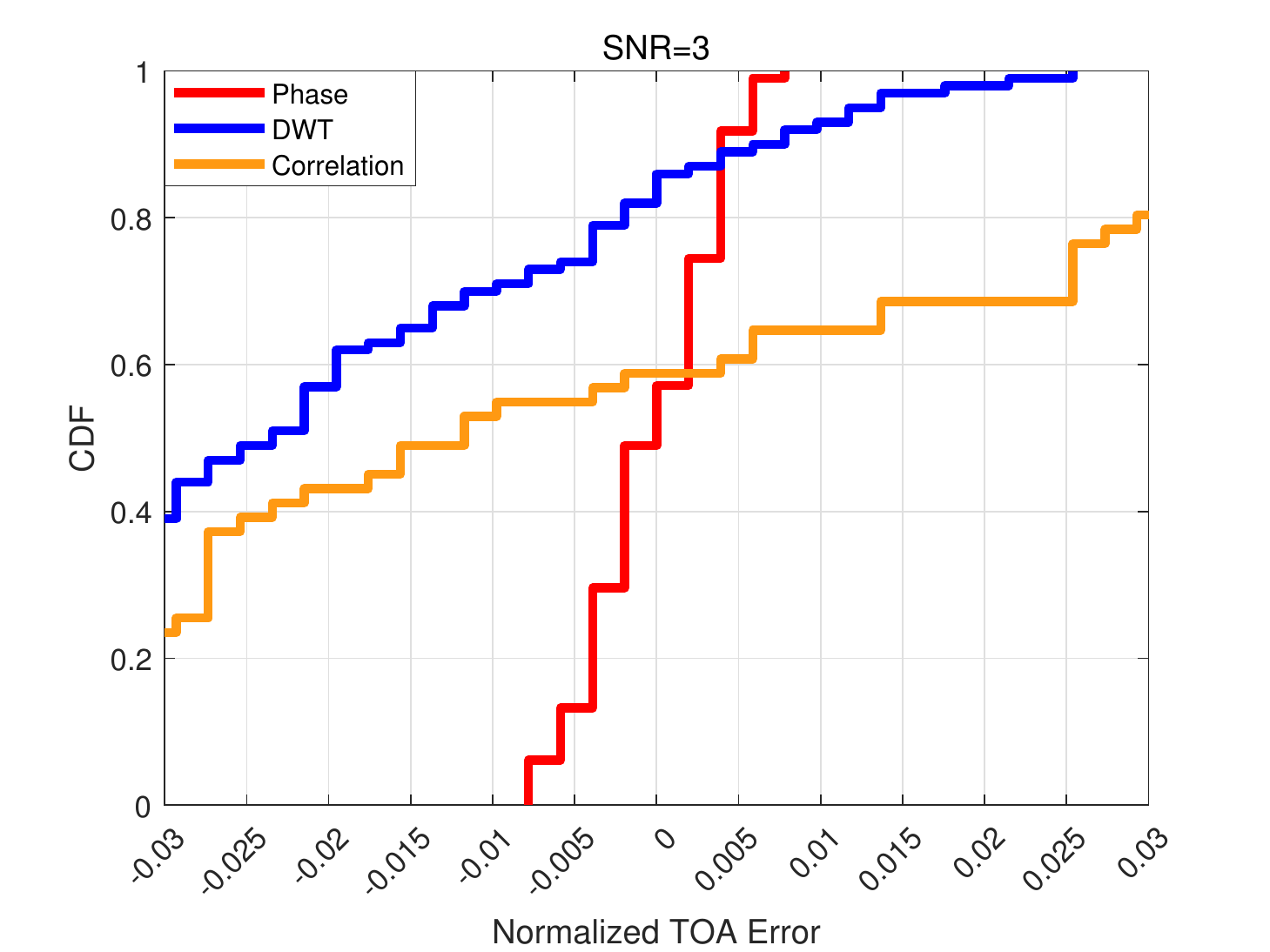}
				\end{minipage}%
		}%
	\subfigure[]{
			\begin{minipage}[t]{0.3\textwidth}
					\centering
					\includegraphics[width=2in]{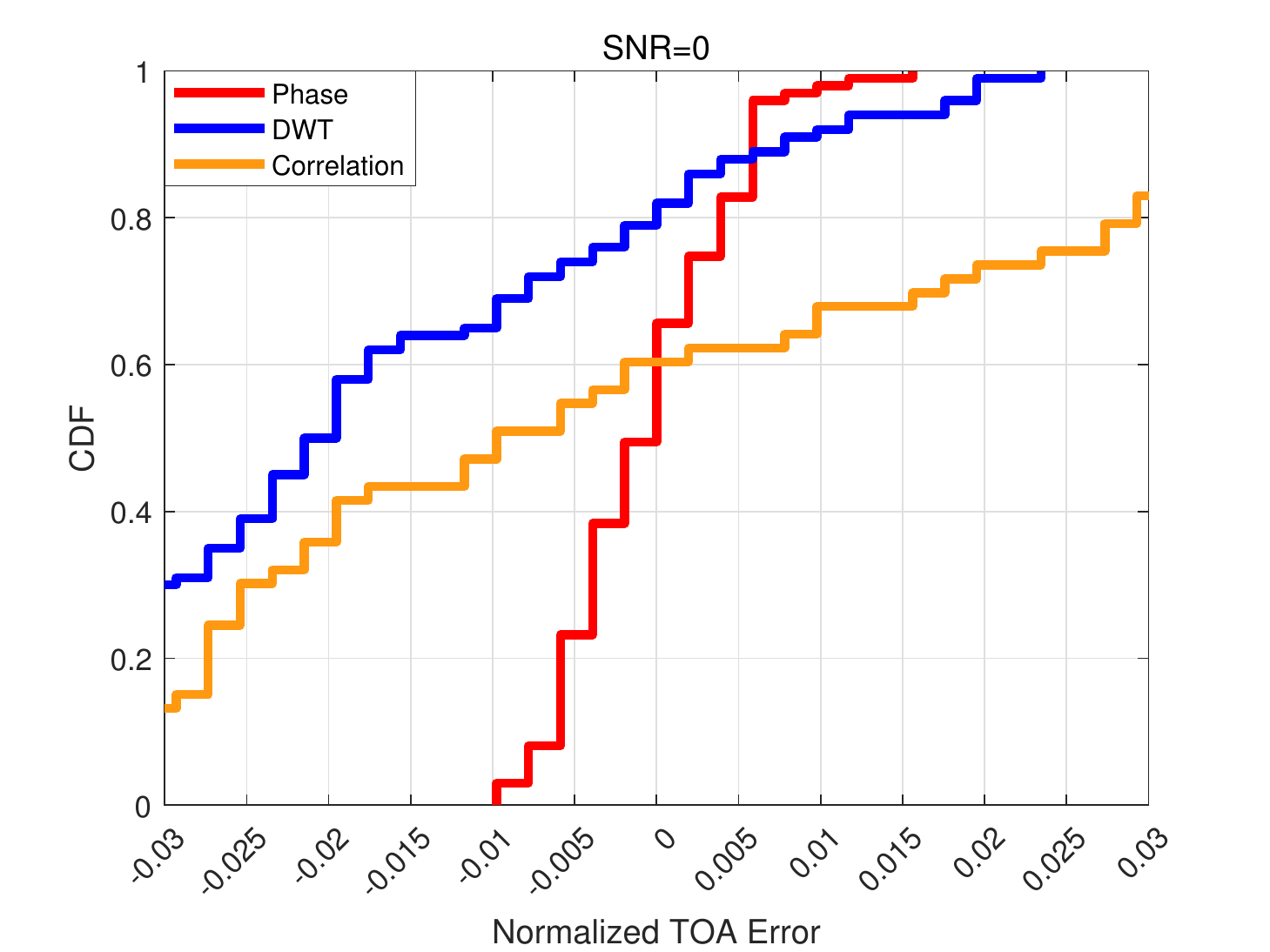}
				\end{minipage}%
		}%
	\subfigure[]{
			\begin{minipage}[t]{0.3\textwidth}
					\centering
					\includegraphics[width=2in]{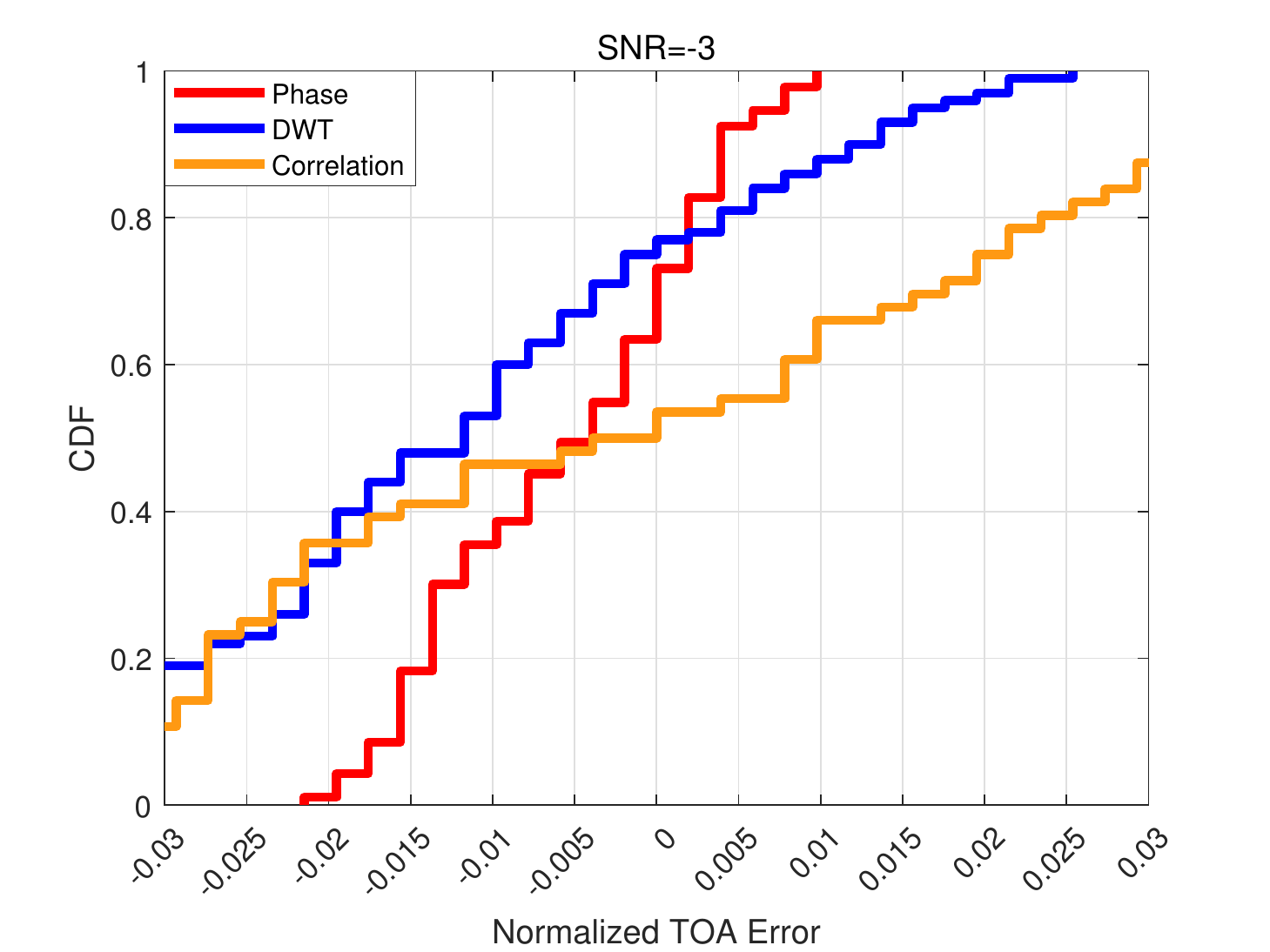}
				\end{minipage}%
		}%
	
	\centering
	\caption{CDF of normalized ToA estimation error on TDL-C channel}
\end{figure*}
\section{conclusion}
In this paper, we propose a phase series based approach using change point detection to address NB-IoT uplink synchronization for NTNs without GNSS. We analyze the linearity of phase series which allows us to identify the expression of ToA and CFO. Then a coarse estimation method to eliminate the impact of large propagation delay is proposed. After compensating the delay, the fine estimation of residual ToA and CFO is presented. Simulation results demonstrate the superiority of our method.

\bibliographystyle{IEEEtran}

\bibliography{ref}
\end{document}